\documentclass[aps,prl,twocolumn,superscriptaddress,amsmath,amssymb]{revtex4}
\usepackage{float}
\usepackage{makeidx}
\usepackage{graphicx,amsmath}
\usepackage{CJK}
\usepackage{colordvi}
\usepackage{color}
\usepackage{amssymb}
\usepackage{epstopdf}
\usepackage{bm}
\usepackage{graphicx,epsfig}
\usepackage{mathrsfs}
\usepackage{dcolumn}
\usepackage{natbib}
\usepackage{miniplot}
\usepackage{subfigure}
\begin{document}
\begin{CJK*}{GB}{gbsn}

\title{Experimental generation of phase wraps for subwavelength phase structures in Bose-Einstein
condensate with two-dimensional optical lattice }

\author{Kai Wen}

\affiliation{State Key Laboratory of Quantum Optics and Quantum
Optics Devices, Institute of Opto-Electronics, Collaborative Innovation Center of Extreme Optics, Shanxi University,
Taiyuan 030006, P.R.China }

\author{Zengming Meng}

\affiliation{State Key Laboratory of Quantum Optics and Quantum
Optics Devices, Institute of Opto-Electronics, Collaborative Innovation Center of Extreme Optics, Shanxi University,
Taiyuan 030006, P.R.China }

\author{Pengjun Wang}

\affiliation{State Key Laboratory of Quantum Optics and Quantum
Optics Devices, Institute of Opto-Electronics, Collaborative Innovation Center of Extreme Optics, Shanxi University,
Taiyuan 030006, P.R.China }

\author{Liangwei Wang}

\affiliation{State Key Laboratory of Quantum Optics and Quantum
Optics Devices, Institute of Opto-Electronics, Collaborative Innovation Center of Extreme Optics, Shanxi University,
Taiyuan 030006, P.R.China }

\author{Liangchao Chen}

\affiliation{State Key Laboratory of Quantum Optics and Quantum
Optics Devices, Institute of Opto-Electronics, Collaborative Innovation Center of Extreme Optics, Shanxi University,
Taiyuan 030006, P.R.China }

\author{Lianghui Huang}

\affiliation{State Key Laboratory of Quantum Optics and Quantum
Optics Devices, Institute of Opto-Electronics, Collaborative Innovation Center of Extreme Optics, Shanxi University,
Taiyuan 030006, P.R.China }

\author{Lihong Zhou}
\affiliation{Beijing National Laboratory for Condensed Matter Physics, Institute of Physics, Chinese Academy of Sciences, Beijing 100190, China}

\author{Xiaoling Cui}
\affiliation{Beijing National Laboratory for Condensed Matter Physics, Institute of Physics, Chinese Academy of Sciences, Beijing 100190, China}
\affiliation{Songshan Lake Materials Laboratory , Dongguan, Guangdong 523808, China}

\author{Jing Zhang$^{\ddagger}$}

\affiliation{State Key Laboratory of Quantum Optics and Quantum Optics Devices,
Institute of Opto-Electronics, Collaborative Innovation Center of Extreme Optics, Shanxi University, Taiyuan 030006, P.R.China }

\begin{abstract}
We report an experimental demonstration of engineering phase wraps for sub-wavelength structure in a Bose-Einstein condensate (BEC) with two-dimensional optical lattices. A short lattice pulse is applied on BEC working in the Kapitza-Dirac (or Raman-Nath) regime, which corresponds to phase modulation imprint on matter wave. When the phase modulation on matter wave is larger than $2\pi$ in a lattice cell, there appears phase wraps with multiple $2\pi$ jumps, generating the sub-wavelength phase structure. The phase wraps for sub-wavelength structure are measured in momentum space via the time-of-flight absorption image, which corresponds to converting phase information into amplitude. 
Moreover, we identify an additional condition for the validity of Kapitza-Dirac regime, which relies crucially on the lattice configurations. This scheme can be used for studying the property of optical lattices and topological defects in matter wave.

\end{abstract}
\maketitle
\end{CJK*}

The resolution of a conventional
optical system is limited by the wavelength of light waves due to diffraction \cite{Rayleigh,Born}.
For a standing wave created by two laser fields
with wavelength $\lambda$, the lattice spacing between adjacent
intensity maxima is $\lambda_{L}/2$, where $\lambda_{L} = \lambda/ sin(\theta/2)$, with $\theta$
being the intersecting angle between the two fields. Therefore, the spatial resolution
of interferometric lithography is always
limited to $\lambda_{L}/2$ \cite{Brueck1998}. Circumventing this limit to create
pattens with spatial resolution smaller than $\lambda_{L}/2$
is not only interesting for a fundamental point of view, but also
relevant for the semiconductor industry. In the past decades, many schemes have been
proposed to improve the spatial resolution of interferometric lithography beyond the diffraction limit,
including the schemes of multiphoton nonlinear processes with classical
lights \cite{Bentley2004,Avi2004,Hemmer2006}, quantum lithography with quantum entanglement \cite{Dowling2000,Kok2001,Agarwal2001,D'Angelo2001},
quantum dark state \cite{Agarwal2006,Kiffner2008,Li2008}, Rabi oscillations \cite{Liao2010,Rui2016}, coherent
atom lithography \cite{Liao2013,Fouda2016}.

\begin{figure}[t]
\centerline{
\includegraphics[width=7cm]{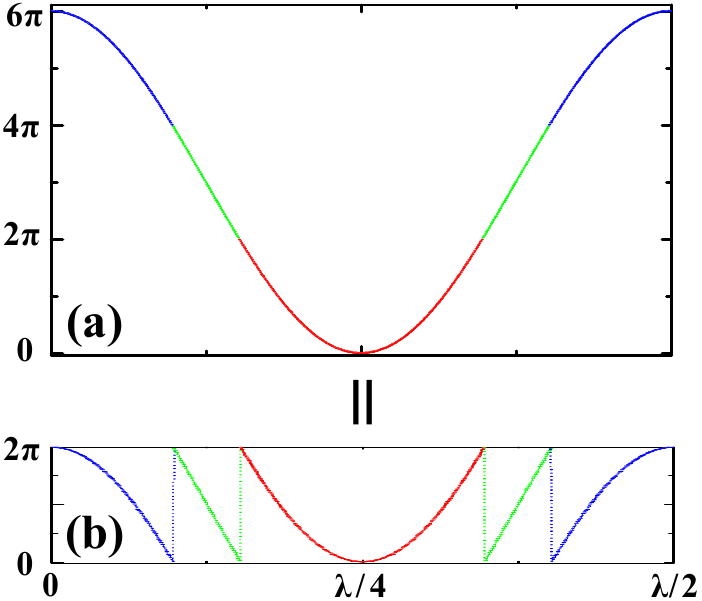}} \vspace{0.1in}
\caption{\textbf{Schematic diagram of the phase wrapping with in a lattice cell.} (a) A continuous phase curve with amplitude larger than $2\pi$. (b) The phase wrap curve with multiple $2\pi$ jumps.
\label{experiment} }
\end{figure}

In this paper, we experimentally develop and demonstrate a new scheme that generate sub-wavelength structure using the idea of phase wrap. In literature, the phase wrapping and unwrapping problems \cite{Itoh1982} have a variety of applications, such as terrain elevation estimation in synthetic aperture radar \cite{Jakowatz2016}, field mapping in magnetic resonance imaging \cite{Fouda1991}, wavefront distortion measurement in adaptive optics \cite{Fried1997}, and is also concerned with topological defects and singularities \cite{Ghiglia1998}.
The idea of phase wrap in our work is implemented in a Bose-Einstein condensate (BEC) by applying two-dimensional (2D) optical lattices. A short pulse of optical lattice is illuminated on BEC working in the regime of Kapitza-Dirac (or Raman-Nath) scattering. In this process, the lattice potential imprints a phase modulation on matter wave in spatial space, as shown schematically in Fig. 1.  When the phase modulation on matter wave is larger than $2\pi$ within a lattice cell, as shown in Fig. 1(a), the continuous phase curve can be equivalent to the phase wrap curve as multiple $2\pi$ jumps forming a saw-tooth waveform in the principal value range of $[0,2\pi]$, as shown in Fig. 1(b). In this case, the phase wrap curve form the sub-wavelength phase structure, and the wrapping number $N$ gives the sub-wavelength $\lambda/(2N)$. We measure sub-wavelength phase structure
in momentum space via the time-of-flight absorption image, which corresponds to converting phase information into amplitude. Here, we work with two different kinds of 2D optical lattices, and the sub-wavelength phase patterns show the properties of the optical lattice.


Now we discuss how the optical lattice pulse imprints a phase modulation on matter wave.
Here we apply a short pulse of optical lattice on BEC working in the Kapitza-Dirac regime \cite{Gould1986,Ovchinnikov1999,Oberthaler1999,Gadway2009} limited to pulse duration $\tau\ll h/4E_{rec}$, where $E_{rec}$ is the photon recoil energy. Previous studies have shown that under a pulse that is short enough, the kinetic energy can be neglected, thus the lattice potential simply acts as a phase grating on the matter wave \cite{Oberthaler1999}. 
In this work, we show that besides the short lattice pulse, the phase grating on matter wave (or the neglecting of kinetic energy) requires an additional condition on the configuration of lattice potentials. Under such condition, the Kapitza-Dirac regime is equivalent to that the effect of short lattice pulse is to impose a spatially periodic phase modulation on the matter wave. 

\begin{figure}
\centerline{
\includegraphics[width=7cm]{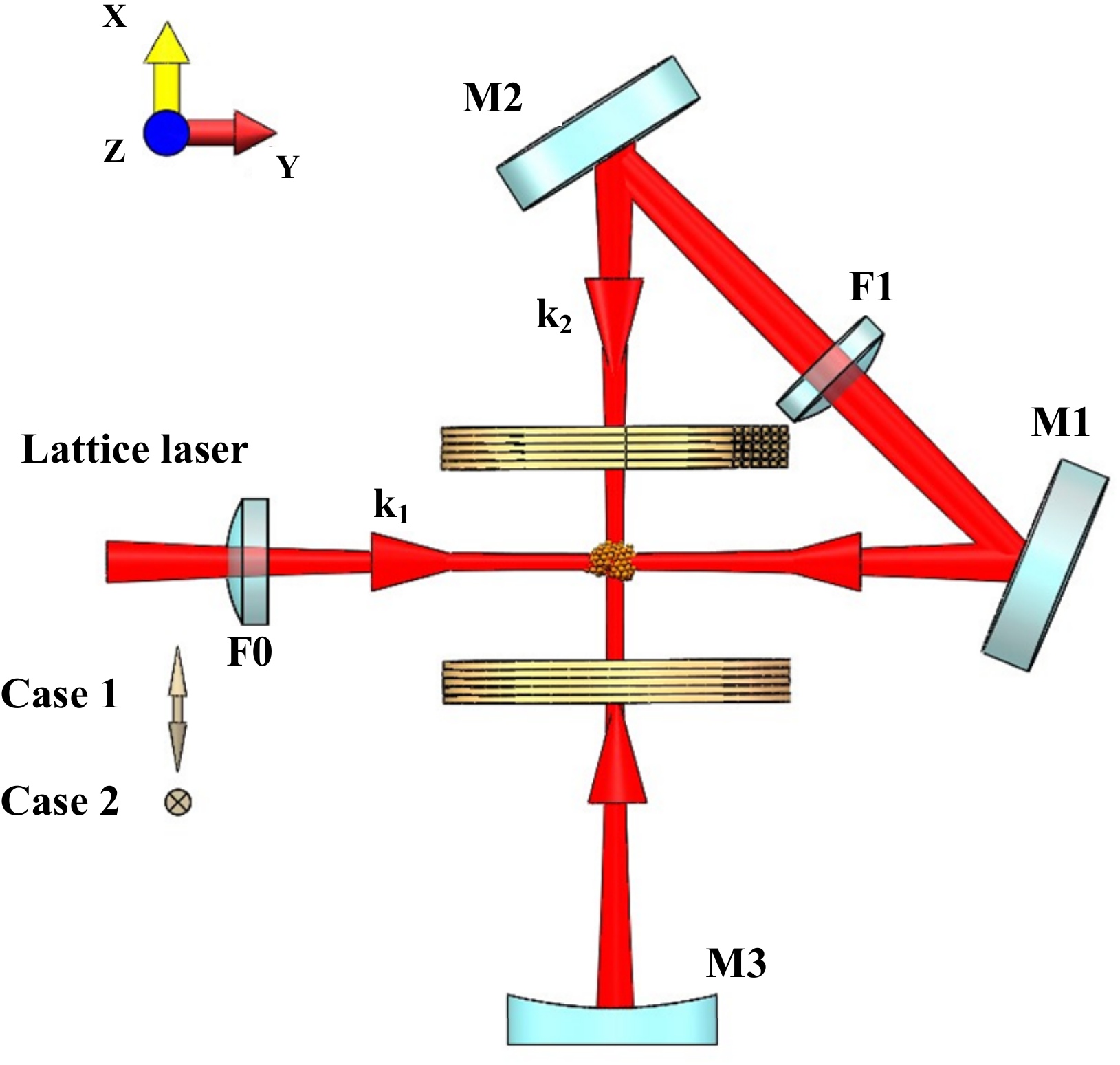}} \vspace{0.1in}
\caption{\textbf{Schematic diagram of the experimental setup to realize the two-dimensional optical lattice.} The two-dimensional optical lattices are made of a single fold retroreflected laser beam. The linear polarization of the incident laser beam aligned normal or parallel to the drawing plane can generate two different kinds of 2D optical lattice potentials.
\label{fig2} }
\end{figure}

In our experiment, the ultracold $^{87}$Rb atoms in the $|F=2,m_{F}=2\rangle$ state is prepared in the crossed optical dipole trap \cite{Xiong2010}. Here $F$ denotes the total angular momentum and $m_{F}$ the magnetic quantum number of the state. Forced evaporation in the optical trap is used to create the BEC with up to $5\times10^5$ atoms, which is used as the coherent matter wave. The lattice beam is derived from a single frequency Ti:sapphire laser and operated at a wavelength of $\lambda=800$ nm. An acousto-optical modulator is used to control the intensity of the lattice beam. Then the light is coupled into a polarization maintaining fiber to provide a clean $TEM_{00}$ spatial mode. A polarizer
after the fiber creates a well defined polarization in the x-z plane. In order to generate two different kinds of 2D optical lattice potentials, we employ the scheme of folded retroreflected mirrors, as shown in Fig. \ref{fig2}. This scheme has been used to experimentally design and implement the 2D optical lattice of double wells suitable for isolating and manipulating an array of individual pairs of atoms \cite{Strabley2006} and predict a topological semimetal in the high orbital bands in this 2D lattice \cite{Sun2012}. The light is folded by plane mirrors M1 and M2
and then retroreflected by concave mirror M3. Therefore, the incoming beam with wave vector $k_{1}$ is reflected by mirrors M1 and M2 and refocus on the atomic cloud with wave vector $k_{2}$, and two wave vectors intersect orthogonally. Lenses F0 and F1 with the concave mirror M3 generate almost the same focus beam radius at the intersection of the four beams with the $1/e^{2}$ radius of 200 $\mu$m.

We first consider the case 1 when the linear polarization of incident laser beam is aligned parallel to the drawing plane (referred as in-plane lattice). Due to the orthogonal intersection of laser beam and the orthogonality of the polarization between $k_{1}$ and $k_{2}$, the resulting 2D lattice is a square lattice formed by two independent 1D lattices. The potential of 2D square lattice is described by
\begin{eqnarray}
U_{1}(x,y)=V[\cos^{2}(kx)+\cos^{2}(ky)]. \label{in_plane}
\end{eqnarray}
Here $k=2\pi/\lambda$ and $\lambda$ denotes the wavelength of the lattice beam. This generates a 2D square lattice with antinode (nodes) spaced by $\lambda/2$ along x and y respectively, 
as shown in Fig. 3(a).

As the case 2, the linear polarization of incident laser beam is aligned normal to the drawing plane (referred as out-plane lattice). In this case the potential is not simply a sum of independent lattices in x and y, but rather given by
\begin{eqnarray}
U_{2}(x,y)=V[\cos(kx)+\cos(ky)]^{2}. \label{out_plane}
\end{eqnarray}
The extra interference term, $2\cos(kx)\cos(ky)$ in Eq. 2, induces the lattice period with $\lambda/\sqrt{2}$ along $x+y$ and $x-y$ directions. When the lattice laser is red detuning ($V<0$), the generated lattice is just like to dug the holes on the ground and there are nodal lines along the diagonals as shown in Fig. 3(c). Different from the square lattice in Eq. 1, here the out-plane lattice in Eq. 2 displays different shapes near the antinode and node. We will show later that this property would significantly affect the sub-wavelength structure as measured in our experiment.

\begin{figure}
\centerline{
\includegraphics[width=7cm]{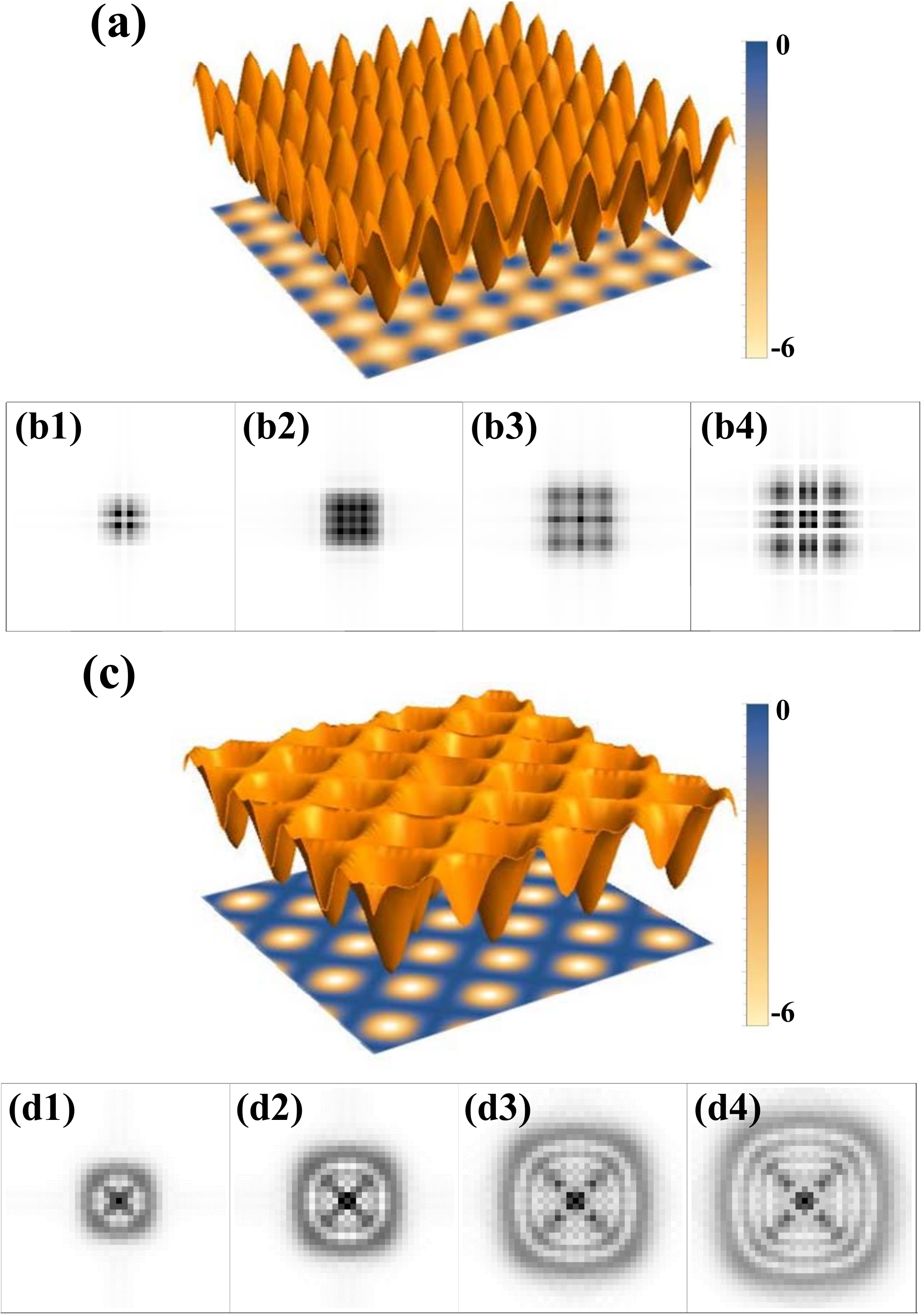}} \vspace{0.1in}
\caption{\textbf{Two types of two-dimensional optical lattices and the associated momentum-space intensity distributions of BEC according to the phase modulation formula. } (a) In-plane lattice (Eq. \ref{in_plane}). (b1)-(b4) Momentum-space intensity distribution of BEC with in-plane lattice pulse for $V\tau/\hbar$=4, 6, 8, 10. (c) Out-plane lattice with red detuning (Eq. \ref{out_plane}). (d1)-(d4) Momentum-space intensity distribution of BEC with out-plane lattice pulse for $|V|\tau/\hbar$=4, 6, 8, 10.
\label{experiment} }
\end{figure}

As discussed previously, when work in the Kapitza-Dirac regime and neglect the kinetic energy of matter wave, the evolution of BEC after a short pulse duration $\tau$ can be described classically by
\begin{eqnarray}
\Psi(x,y)=A e^{iU(x,y)\tau/\hbar}. \label{psi}
\end{eqnarray}
Here $A$ is the Gaussian wave function of initial state of BEC, while the lattice potential $U(x,y)$ imprints a phase modulation on the BEC. 
Take the in-plane lattice for example, when $|V|\tau/\hbar>\pi$ the phase of BEC within a unit cell appears as multiple $2\pi$ jumps, known as phase wraps. Similarly, the phase wrap occurs for out-plane lattice when $|V|\tau/\hbar>\pi/2$. Such phase wraps equivalently generate the sub-wavelength structure, as schematically shown in Fig. 1, where the wrapping number $N$ gives the sub-wavelength $\lambda/(2N)$.
Since we can not directly measure the phase structure of BEC by the in-situ imaging, we must convert the phase information into amplitude. The easiest way to realize this goal is to take the Fourier transform
\begin{eqnarray}
\Psi(p_x,p_y)=F[\Psi(x,y)].  \label{psi_p}
\end{eqnarray}
This transformation converts the phase information of real-space wave function to the amplitude information in momentum space. The resulted intensity distribution $I_{p_x,p_y}=|\Psi(p_x,p_y)|^{2}$ in momentum space can be directly measured in
the ultracold atomic experiment through the time-of-flight absorption image. 

In Figs. 3(b1)-(b4) and (d1)-(d4), we show the intensity distributions $I_{p_x,p_y}$ from Eqs. \ref{psi} and \ref{psi_p} corresponding to the two different lattice potentials as displayed, respectively, in Fig. 3(a) and (c). Starting with the initial BEC which has a single momentum component ($k=0$), the phase modulations with the 2D optical lattices generate discrete momentum components. Remarkably, these discrete components organize themselves into different pattens in a larger momentum scale, which is related to phase wraps in a smaller length scale (within a single cell), leading to the sub-wavelength structure. Since the in-plane lattice generates the sub-wavelength phase structure with the trenches along x and y (Fig. 3(a)), the intensity distribution after the Fourier transform presents the larger line structures along x and y, see Fig. 3(b1)-(b4). In contrast, out-plane lattice generates the sub-wavelength phase structure with ring structure (Fig. 3(c)) and the intensity distribution after the Fourier transform presents the larger ring structures, see Fig. 3(d1)-(d4).


\begin{figure}
\centerline{
\includegraphics[width=8cm]{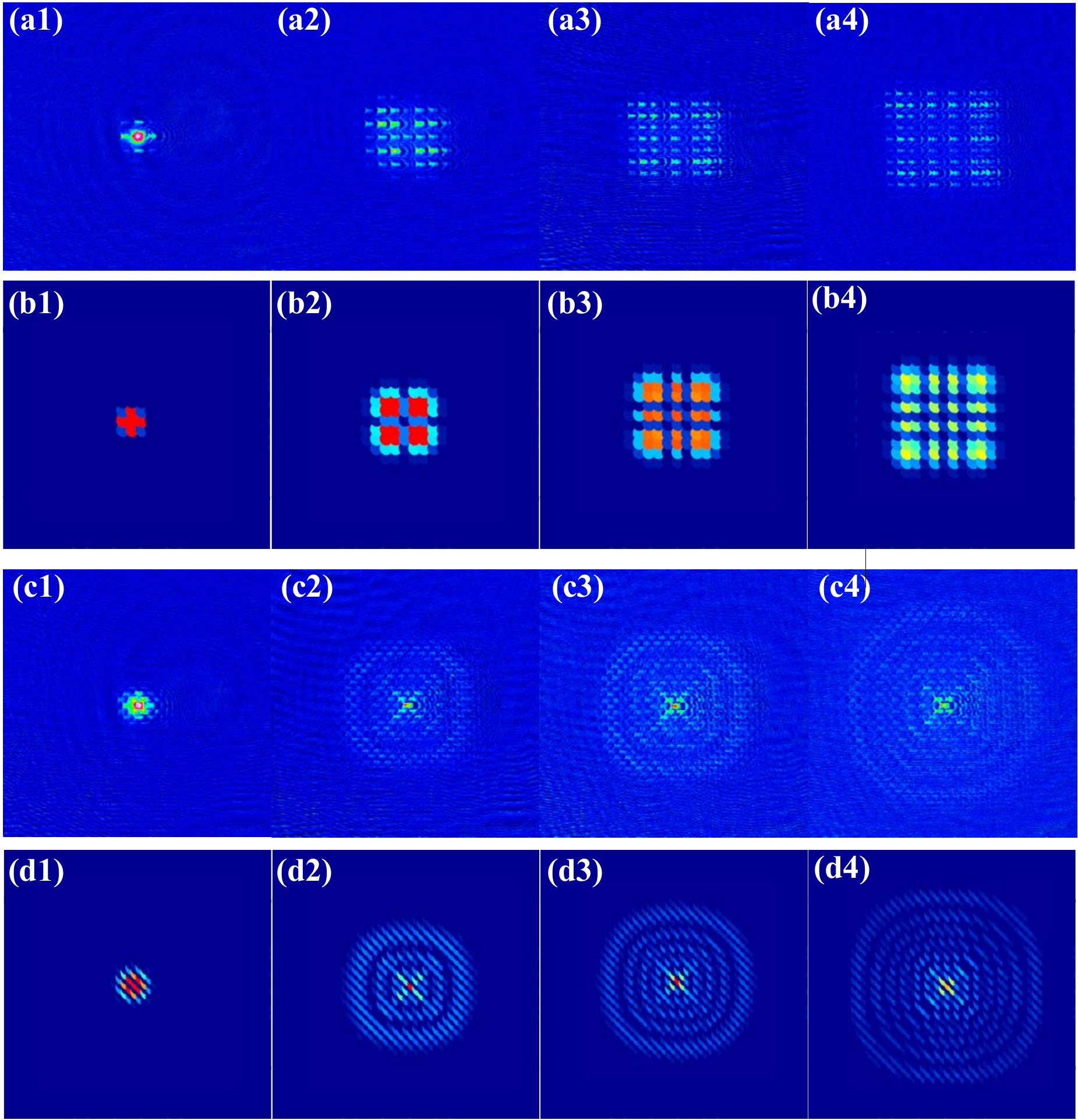}} \vspace{0.1in}
\caption{\textbf{Atomic density distribution of the time-of-flight absorption image after applying 2D optical lattice pulse on BEC for red detuning ($V<0$).} (a1)-(a4) Experimental data for in-plane lattice pulse with $|V|/E_{rec}=$ 10, 59, 88, 123, in comparison to the theoretical results from full quantum evolutions (b1)-(b4). 
(c1)-(c4) Experimental data for out-plane lattice pulse with $|V|/E_{rec}=$ 10, 59, 88, 123, in comparison to the theoretical results from full quantum evolutions (d1)-(d4) .
The lattice wavelength is $\lambda=800$ nm. The duration time of applying 2D optical lattice pulse on BEC is 4 $\mu s$ and TOF=7 ms.
\label{experiment} }
\end{figure}

In the experiment, $h/4E_{rec}=70$ $\mu s$ for the lattice wavelength of $\lambda=800$ nm, which is red detuning corresponding to $V<0$ in Eqs. 1 and 2 . We apply 2D optical lattice short pulse with 4 $\mu s$ on BEC, which work in the Kapitza-Dirac regime. Then we immediately turn off the optical trap, let the atoms ballistically expand in 7 ms and take the absorption image. Fig. 4(a1)-(a4) show the atomic density distribution of the time-of-flight absorption image after applying in-plane lattice pulse on BEC and Fig. 4(c1)-(c4) are for out-plane lattice. These experimental data are in good consistence with theoretical results from the full quantum evolution of the BEC (see supplementary material), as shown by Figs. 4(b1)-(b4) and 4(d1)-(d4). The results are also qualitatively consistent with those from the classical treatment (Eq. 3) (see supplementary material \cite{sup}).

As expected, the atomic density distributions of the time-of-flight absorption image exhibit discrete momentum components. 
Nevertheless, the distribution of the discrete momentum components is not uniform, which depends on the sub-wavelength phase structure in a lattice cell. In a larger momentum scale, the distribution of discrete momentum components shows line structure along x and y for in-plane lattice, and ring structure for out-plane lattice. 
As increasing the lattice laser power, more and more lines or rings appear, see Fig. 4. This can be well explained by the phase wrap picture since more and more $2\pi$ phase jumps occurs within the single cell as increasing $|V|$, giving more delicate structure with smaller sub-wavelength. In this way, the line or ring pattens in a large momentum scale of the intensity distribution directly reflect the sub-wavelength structure of particular lattice potentials.


\begin{figure}
\centerline{
\includegraphics[width=9cm]{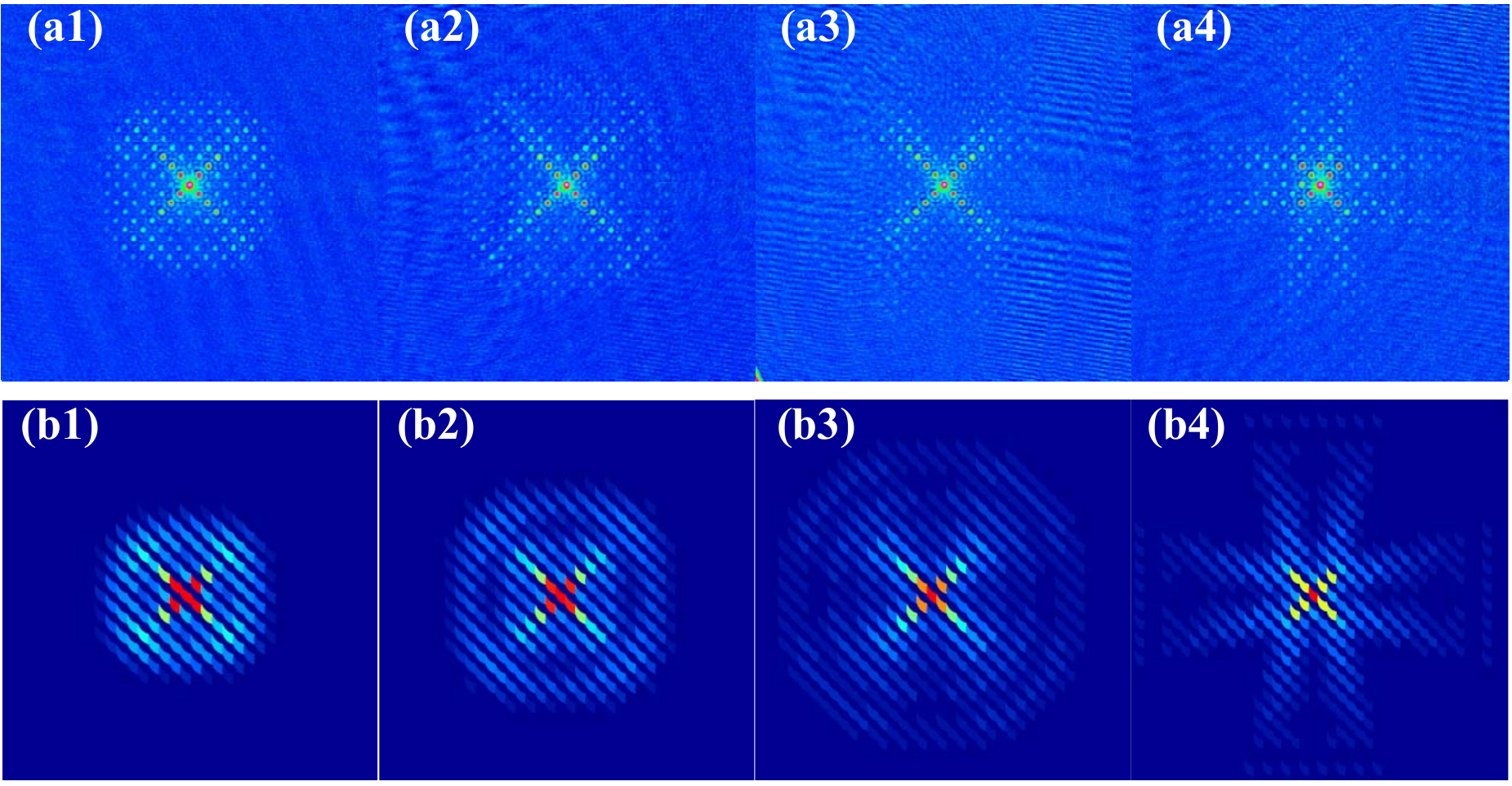}} \vspace{0.1in}
\caption{
\textbf{Atomic density distribution of the time-of-flight absorption image after applying 2D out-plane lattice with blue detuning.} (a1)-(a4) Experimental data with $V/E_{rec}=$34, 53, 82, 120, in comparison to the theoretical results from full quantum evolutions (b1)-(b4).
The lattice wavelength is $\lambda=793.4$ nm. The duration time of applying 2D optical lattice pulse on BEC is 4 $\mu s$ and TOF=7 ms .
\label{experiment} }
\end{figure}


To this end, we have shown the validity of classical treatment (Eq. 3) in predicting the BEC dynamics, which neglects the quantum motion of the wave-packet due to the kinetic term. Nevertheless, it is noted that the classical treatment does not only reply on the short lattice pulse, but also require an additional condition on the configuration of lattice potentials. In above experiments, we consider lattice potentials $U_1$ and $U_2$ in Eqs. 1 and 2 exhibiting sharp structure near the bottom, as seen from Fig. 3(a) and (c). In this case, the low-energy states, which have the largest overlap with the initial BEC wave-packet and contribute most to the BEC dynamics, are well localized and thus able to reflect the structure of lattice potentials. As a result, the phase modulation formula (Eq. 3) can be a good approximation to mimic the BEC dynamics. To summarize, this additional condition is that the lattice supported low-energy states are sufficiently localized, such that the quantum motion of BEC can be efficiently suppressed to allow the neglecting of kinetic energy.


To further demonstrate this additional condition, we have performed an additional set of experiments with the out-of-plane lattice pulse switching to blue detuning, i.e., with  $V>0$ in Eq. 2. In this case, the bottom of the lattice potential is fairly smooth and the quantum motion of low-energy states can not be neglected, which does not satisfy the additional condition of neglecting the kinetic energy. The phase modulation formula Eq. 3 is not applicable any more. Experimentally, no clear ring structure can be found in large $V$ limit, see Fig. 5, in contrast to the red detuning case with large $|V|$. The observation is consistent with results from the full quantum calculation but not with the classical treatment, see supplementary material \cite{sup}.

In conclusion, we have experimentally observed the phase wraps for sub-wavelength structure in BEC with two-dimensional optical lattices. The phase wraps are realized by applying  a short lattice pulse in the Kapitza-Dirac regime, such that the lattice potentials imprint the phase modulation on matter wave. We have detected this sub-wavelength phase structure in a lattice cell by measuring the intensity distribution of BEC in momentum space, which shows the line or ring structures in a larger momentum scale. Furthermore, we have demonstrated an additional condition for the application of the Kapitza-Dirac regime where the kinetic energy can be neglected. The result can be used to detect more properties of the optical lattice, such as the shape of the lattice cell and the localized property of low-energy Bloch states. Finally, the phase wrapping picture in our work can also be connected to the topological defects in matter wave, such as phase steps or vortices.

$^{\ddagger}$Corresponding author email: jzhang74@sxu.edu.cn,
jzhang74@yahoo.com.

\begin{acknowledgments}
This research is supported by the National Key Research and Development Program of China (2016YFA0301602, 2018YFA0307600, 2016YFA0300603),
NSFC (Grant No. 11234008, 11474188, 11704234, No.11622436, No.11421092, No.11534014) and the Fund for
Shanxi "1331 Project" Key Subjects Construction.

\end{acknowledgments}

\end{document}